\documentclass[12pt]{iopart}
\usepackage{iopams,graphicx}

\usepackage{lmodern}
\usepackage[T1]{fontenc}
\usepackage{bbold}
\graphicspath{{Graphics/}}
\usepackage{xcolor}

\usepackage{soul}
\usepackage[mathscr]{euscript}
\usepackage{cite}
\usepackage[super]{nth}
\usepackage{float}  
\usepackage{comment} 
\usepackage[colorlinks, citecolor = blue, urlcolor = blue]{hyperref}

\begin{document}

\title{Freezing transitions of Brownian particles in confining potentials}

\author{Gabriel Mercado-V\'asquez$^1$, Denis Boyer$^1$,	and Satya N. Majumdar$^2$}

\address{$^1$ Instituto de F\'isica, Universidad Nacional Aut\'onoma de M\'exico, Mexico City 04510, Mexico}
\address{$^2$ LPTMS, CNRS, Univ. Paris-Sud, Universit\'e Paris-Saclay, 91405 Orsay, France}


\begin{abstract}
We study the mean first passage time (MFPT) to an absorbing target of a one-dimensional Brownian particle subject to an external potential $v(x)$ in a finite domain. We focus on the cases in which the external potential is confining, of the form $v(x)=k|x-x_0|^n/n$, and where the particle's initial position coincides with $x_0$. We first consider a particle between an absorbing target at $x=0$ and a reflective wall at $x=c$. At fixed $x_0$, we show that when the target distance $c$ exceeds a critical value, there exists a nonzero  optimal stiffness $k_{\rm opt}$ that minimizes the MFPT to the target. However, when $c$ lies below the critical value, the optimal stiffness $k_{\rm opt}$ vanishes. Hence, for any value of $n$,  the optimal potential stiffness undergoes a continuous \lq\lq freezing\rq\rq\ transition as the domain size is varied. On the other hand, when the reflective wall is replaced by a second absorbing target, the freezing transition in $k_{\rm opt}$ becomes discontinuous. The phase diagram in the $(x_0,n)$-plane then exhibits three dynamical phases and metastability, with a \lq\lq triple\rq\rq\  point at $(x_0/c\simeq 0.17185$, $n\simeq 0.39539)$. For harmonic or higher order potentials $(n\ge 2)$, the MFPT always increases with $k$ at small $k$, for any $x_0$ or domain size. These results are contrasted with problems of diffusion under optimal resetting in bounded domains.  
\end{abstract}


\maketitle

\section{Introduction}

The theory of first passage processes is concerned with the statistical properties of the time it takes for a stochastic process to reach a specific configuration or site for the first time \cite{van1992stochastic,redner2001guide,bray2013persistence}. The first passage time (FPT) is of primary relevance in problems across many areas of science, ranging from chemical reaction kinetics \cite{Szabo1980,cui2006mean,zhang2016first,chou2014first}, neuronal science\cite{gerstein1964random,beggs2003neuronal}, animal foraging\cite{viswanathan2011physics,kagan2015search}, or finance\cite{chicheportiche2014some}. In many situations, the motion of a diffusive entity is restricted by the surroundings of the medium or some particular limits. The introduction of boundaries has many implications on the particle dynamics and the statistical properties of the FPT \cite{redner2001guide,benichou2010geometry}. For simple diffusive processes, it is well-known that the mean first passage time (MFPT) to an absorbing target is finite in a bounded domain, unlike in the open geometry where it is infinite\cite{redner2001guide}.  At the biological level the presence of boundaries can influence the completion of vital processes: gene transcription is strongly controlled by the geometry of the medium; in crowded environments composed of many obstacles, the time at which a gene is activated is extremely sensitive to changes in the initial position of the transcription factors \cite{benichou2010geometry,BENICHOU2014225}. To achieve their biological function, certain substances must diffuse inside the cell volume until they escape from the confinement through a narrow window\cite{BenichouNarrow2008}. Red blood cells are oxygenated once the oxygen reaches the cell wall and diffuses within the cell interior\cite{bamford1985diffusion}. The presence of  boundaries can also affect reaction kinetics when there are more than one target site \cite{Chevalier_2010}. 


Placing hard obstacles in the bulk of the system is not the only way to restrict the motion of particles; external forces can also serve for this purpose.  In the celebrated Kramers problem\cite{kramers1940brownian,KramersHistory1990}, a Brownian particle is trapped in a potential well; to reach another (deeper) well the particle must surpass a potential barrier with the assistance of thermal fluctuations. 
In a polymer chain, for instance, the two ends of the chain have to overcome the harmonic potential between them in order to react\cite{Szabo1980}. Potential barriers strongly influence the first passage statistics of diffusive systems. In particular, the shape of the potential can have unexpected consequences on the rate at which the FPT probability density decays at large times\cite{sabhapandit2020freezing}.

As the efficiency of a search process can be measured through the MFPT, an important question one may ask is: which shape should an external potential have in order to minimize the time taken by a particle to encounter its target? This question has been addressed  in different contexts. For instance, a piece-wise linear potential fluctuating in time between two different barrier heights produces a stochastic resonant activation that can lower the MFPT compared to the time taken with the average potential \cite{ResonantDoering1992}. In the context of static potentials, the introduction of a high but narrow barrier near a hard wall markedly reduces the MFPT to a target site located on the other side of the barrier \cite{Palyulin_2012,Chupeau2020}. Extensions have also considered subdiffusive processes \cite{Palyulin_2013}.
When the target is randomly placed at a position chosen from a given distribution, the optimal potential shape can be deduced from a variational calculation \cite{Ku_mierz_2017}. 

In recent years, it has been shown that random searches in equilibrium can be further optimized by driving the searcher out of equilibrium. One way of doing this is for instance through resetting, {\it i.e.}, by means of stochastic interruptions of the motion that bring the searcher back to a particular position of space, from which movement starts afresh
\cite{EvansPRL2011}. These interruptions can benefit the search process by cutting off those fruitless trajectories that move away from the target position. The MFPT can be minimized with respect to the resetting rate for a variety of background processes, such as Brownian motion \cite{Evans_2011,Evans_2020} or L\'evy flights\cite{kusmierz2014first}. Stochastic processes with resetting also exhibit interesting features such as the emergence of non-equilibrium steady states (NESS) \cite{manrubia1999stochastic,Evans_2011,evans2018effects}, whose relaxation dynamics is quite peculiar \cite{majumdar2015dynamical,Gupta_2021,Singh_2020}. 


The stationary distribution of a Brownian motion with diffusion constant $D$ under resetting at finite rate $r$ resembles a Boltzmann-Gibbs steady state with an effective potential of the form $k|x-x_0|$, where $x_0$ is the resetting position and $k=\sqrt{r/D}$ \cite{Evans_2013}.  However, motion under resetting is markedly different from a Langevin dynamics in this effective potential. On a semi-infinite line with an absorbing wall located at the origin, it is also possible to minimize with respect to $k$ the MFPT given by the Kramer's theory and the above  piece-wise linear potential. It is found that the MFPT of the particle under resetting at the optimal rate is lower than the optimal MFPT in the Kramers case \cite{Evans_2013,giuggioli2019comparison}. 

Nevertheless, in finite domains, resetting does not always represent an efficient search strategy but can be detrimental instead. Depending on the location of the resetting position with respect to the boundaries, resetting can either decrease or increase the MFPT that would be obtained with simple diffusion \cite{Christou_2015,Pal2019}. As $x_0$ is varied away from the target, the optimal resetting rate $r_{\rm opt}$ typically undergoes a continuous "freezing" transition from $r_{\rm opt}>0$ to $r_{\rm opt}=0$ (equivalent to free diffusion) at a particular critical position $x_c$. If the diffusive particle is additionally subject to an external potential, the interplay between resetting and the crossing of potential barriers (in finite or infinite intervals) can also produce transitions where the advantage of resetting vanishes \cite{singh2020resetting,ray2021resetting,ahmad2022first}. In situations where the potential attracts the particle toward the target, strong enough potentials can also suppress the benefit of resetting to the initial position \cite{Saeed2019,ahmad2022first}. In those cases, there are ranges of values of the potential strength for which $r_{\rm opt}$ vanishes.
These intriguing transitions have been studied for resetting problems, sometimes in combination with external potentials as mentioned above. However, to our knowledge, a systematic analysis of random search optimization in finite intervals using external potentials only (without resetting) is still lacking.

\begin{figure}[htp]
\centering
			\includegraphics[width=.8\textwidth]{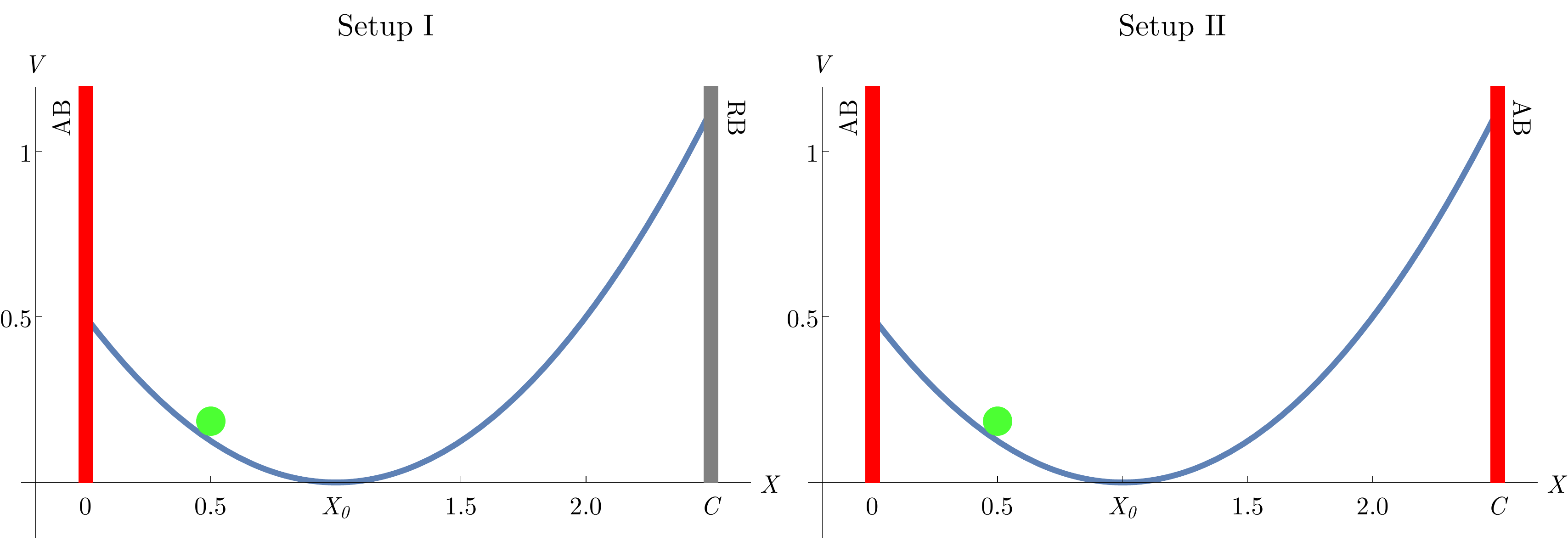}
			\caption{The two set-ups investigated here. A Brownian particle (green point) is subject to a potential $v(x)$ with a minimum at $x_0=1$. In set-up I, an absorbing boundary (AB) is placed at the origin, whereas a reflective boundary (RB) is placed at $c$. In set-up II, both walls are absorbing.  }
			\label{fig:fig1ab}
\end{figure}

We aim in this work to study target encounter optimization in the presence of confining potentials in finite domains, and to contrast with the results that are known for resetting problems in similar geometries. The motion of the particle is confined within a one-dimensional interval $[0,c]$ in two different set-ups: In set-up I (see Fig. \ref{fig:fig1ab}-left), an absorbing boundary is located at the origin whereas a reflective wall is placed at $x=c$. In set-up II (Fig. \ref{fig:fig1ab}-right), both boundaries are absorbing. We consider here external potentials that are minimum at a position $x_0$, of the form $v(x)=k|x-x_0|^n/n$, where $n>0$, $k\ge 0$. The limiting case $n\rightarrow 0$ corresponds to the logarithmic potential. We are mainly interested in the effects that the parameters $k$, $x_0$ and $n$ have on the mean first passage time properties of the system. Although this problem is directly related to the classical Kramers' problem, it exhibits rich behaviours that are not observed in unbounded domains, or even for free diffusion with stochastic resetting. 

This article is organized as follows: In Section \ref{generalsetup} we compute the mean time needed for a Brownian particle subject to a potential $v(x)$ to reach a target with the two different boundary conditions. Section \ref{refboundary} is devoted to the investigation of set-up I. We analyze the expression of the MFPT for a particle starting from $x_0$  as a function of the potential stiffness $k$, finding that, when $x_0/c$ is below a critical value, the MFPT can be minimized at a non-zero stiffness $k_{\rm opt}$, whereas above the critical value, the mean search time is optimized by setting $k=0$. In section \ref{secResetting}, we compare this "freezing" transition for $k_{\rm opt}$, which is continuous, to the similar transition that occurs in this geometry for the problem of free diffusion under resetting. In Section \ref{twotargets} we compute and analyze the MFPT for the set-up II (or the mean exit time of the particle from the interval $[0,c]$). We observe qualitatively different behaviours: due to metastability effects, the transition in $k_{\rm opt}$ at fixed $n$ is discontinuous as $x_0/c$ crosses a transition point, at least for $n>0.39539...$. Finally, we discuss our findings in Section \ref{discussion}.


\section{General set-up
and the solution}\label{generalsetup}

In this section we shall recall some well-known results in the literature \cite{risken1996fokker, gardiner2004handbook}. This will help set up our notations and the analysis to follow in subsequent sections. 
The evolution of a Brownian particle with position $X(t)$ in a potential $V(X)$ is given by the over-damped Langevin equation:
\begin{equation}
    \frac{dX(t)}{dt}= -\frac{V^{\prime}(X(t))}{\Gamma}+\sqrt{2D}\xi(t),
    \label{ruleX1}
\end{equation}
where $\Gamma$ is the friction coefficient,   $\xi(t)$ a Gaussian white noise of zero mean and correlations $\langle \xi(t)\xi(t')\rangle=\delta(t-t')$, and $D$ the diffusion coefficient. The particle is confined in a domain of size  $C$, or $0\le X\le C$.

In the following it is convenient to introduce the dimensionless space and time variables $x=X/L$ and $t/(L^2/D)$ (which we re-note as $t$), where $L$ is an arbitrary length. If $V(X)$ has a single minimum, $L$ can be chosen as the distance $X_0$ between the potential minimum and the target placed at $X=0$. Alternatively, $L$ can be set to the domain size $C$. The re-scaled potential is given by $v(x)=V(xL)/(\Gamma D)=V(xL)/(k_BT)$. The re-scaled domain size, potential minimum position and particle's initial position are denoted as $c=C/L$, $x_0=X_0/L$ and $x_i=X_i/L$, respectively.  In these dimensionless variables, the Langevin equation (\ref{ruleX1}) reduces to
\begin{equation}
\frac{dx}{dt}= - v'(x) + \sqrt{2} \eta(t)
\end{equation}
where $\langle \eta(t)\rangle=0$ and $\langle \eta(t)\eta(t')\rangle= \delta(t-t')$.

We define the survival probability $Q(x_i,t)$ as the probability that the diffusing particle has not been absorbed yet at time $t$, given the initial position $x_i$. We re-note $x_i=x$ in the following. This probability satisfies the backward Fokker-Planck equation (see {\it e.g.} \cite{risken1996fokker})
\begin{equation}
    \frac{\partial Q(x,t)}{\partial t}=\frac{\partial^2 Q(x,t)}{\partial x^2}-v^{\prime}(x)\frac{\partial Q(x,t)}{\partial x}.\label{SurvP}
\end{equation}
Defining the Laplace transform  $\widetilde{Q}(x,s)=\int^{\infty}_0 dt\  e^{-st}Q(x,t)$ and integrating by parts, Eq. (\ref{SurvP}) can be recast as
\begin{equation}
    \frac{\partial^2 \widetilde{Q}(x,s)}{\partial x^2}-v^{\prime}(x)\frac{\partial \widetilde{Q}(x,s)}{\partial x}-s\widetilde{Q}(x,s)=-1,\label{SurvPs}
\end{equation}
where we have imposed the initial condition $Q(x,t=0)=1$ for all $0<x<c$.

As the motion of the particle is confined by the potential, the problem allows us to compute the mean first passage time (MFPT), which is the mean time at which the particle reaches an absorbing boundary for the first time. This quantity can be easily computed knowing the Laplace transform of the survival probability through the usual relation
\begin{equation}
    T(x)=\int_0^\infty Q(x,t) dt=\widetilde{Q}(x,s=0).\label{relQT}
\end{equation}
Therefore, taking $s=0$ in Eq. (\ref{SurvPs}) we obtain the equation for the MFPT $T(x)$ for a Brownian particle in the presence of the potential $v(x)$ 
\begin{equation}
    \frac{\partial^2 T(x)}{\partial x^2}-v^{\prime}(x)\frac{\partial T(x)}{\partial x}=-1,\label{MFPT}
\end{equation}
whose general solution is obtained by direct integration as
\begin{equation}
    T(x)=A\int^x_0 dy\ e^{v(y)}+B-\int^x_0 dy \int^y_0 dz\ e^{v(y)-v(z)}.\label{MFPTsolgral}
\end{equation}
The first two terms solve the homogeneous equation whereas the last one is the particular inhomogeneous solution. The constants of integration $A$ and $B$ are determined by the boundary conditions. In the two set-ups considered here, an absorbing target is placed at the origin, therefore
\begin{equation}
T(x=0)=0,\label{BCmfpt1}
\end{equation}
as a particle starting at $x=0$ is immediately absorbed. Hence, we have $B=0$ in Eq. (\ref{MFPTsolgral}), or
\begin{equation}
    T(x)=A\int^x_0 dy\ e^{v(y)}-\int^x_0 dy\int^y_0 dz\ e^{v(y)-v(z)}.\label{generalmfpt}
\end{equation}

In set-up I, a reflective wall is placed at $x=c$, 
which implies
\begin{equation}
    \frac{\partial T(x)}{\partial x}\Big|_{x=c}=0.\label{BCmfpt2}
\end{equation}
This boundary condition gives, from  Eq. (\ref{generalmfpt}), 
\begin{equation}
A_I=\int^c_0 dz\ e^{-v(z)}    
\end{equation}
and the MFPT for set-up I,
\begin{equation}
    T_I(x)=\int^x_{0}dy \int^c_ydz\ e^{v(y)-v(z)},\label{mfpt2}
\end{equation}
a result obtained in \cite{Szabo1980}, for instance.

In set-up II, another absorbing target is placed at $x=c$ and the boundary condition becomes 
\begin{equation}
    T(x=c)=0.\label{BC2mfpt2}
\end{equation}
From Eq. (\ref{generalmfpt}), the constant is now given by 
\begin{equation}
A_{II}=\frac{\int^c_{0}dy \int^y_0dz\ e^{v(y)-v(z)}}{\int^c_0 dy\ e^{v(y)}}    
\end{equation}
and the solution by
\begin{equation}
    T_{II}(x)=\frac{\int^c_{0}dy \int^y_0dz\ e^{v(y)-v(z)}}{\int^c_0 dy\ e^{v(y)}} \int^x_0 dy\ e^{v(y)} -\int^x_{0}dy \int^y_0dz\ e^{v(y)-v(z)},\label{2mfpt}
\end{equation}
as obtained in \cite{gardiner2004handbook}, for instance.

In the following we consider a potential of the form $v(x)=k|x-x_0|^n/n$, where $n>0$ and $k\ge 0$, {\it i.e.}, with a minimum at $x_0$. The dimensionless parameter $k$ is related to the dimensional stiffness $K$ through the expression $k=KL^n/(k_{B}T)$. With this potential, the general expression of the MFPT in Eq. (\ref{generalmfpt}) can be recast as
\begin{equation}
    T(x)=A\int^x_0 dy\ e^{\frac{k}{n}|y-x_0|^n}-\int^x_0 dy\int^y_0 dz\ e^{\frac{k}{n}\left(|y-x_0|^n-|z-x_0|^n\right)}\label{generalmfptVn},
\end{equation}
where $A$ takes the value
\begin{equation}
    A_{I}=\int^c_0 dz\ e^{-\frac{k}{n}|z-x_0|^n}  \label{A1n} 
\end{equation}
with the reflective boundary at $c$, or
\begin{equation}
A_{II}=\frac{\int^c_{0}dy \int^y_0dz\ e^{\frac{k}{n}\left(|y-x_0|^n-|z-x_0|^n\right)}}{\int^c_0 dy\ e^{\frac{k}{n}|y-x_0|^n}}    \label{A2n}
\end{equation}
in the case of an absorbing boundary at $c$.

\section{Set-up I: absorbing/reflective boundaries}\label{refboundary}
In this section we choose $L=X_0$ as the characteristic length of the system, hence, the dimensionless space variable is $x=X/X_0$ and $x_0=1$. In order to analyze the behaviour of the MFPT as a function of the potential order $n$ and its stiffness $k$, we consider the case in which the initial position of the particle coincides with the minimum of the potential ($x=1$).
In this case, the MFPT will be simply denoted as $T_I$ and from Eqs. (\ref{generalmfptVn}) and (\ref{A1n}), it can be written as 
\begin{equation}
    T_I=\int^1_{0}dy \int^c_ydz\ e^{\frac{k}{n}\left(|y-1|^n-|z-1|^n\right)}.\label{mfptVn}
\end{equation}
The reflective wall is obviously placed further to the right side of the initial position, therefore $c\ge 1$. 

\subsection{Optimal potential stiffness \texorpdfstring{$k$}{Lg}}\label{optimalkreflect}

Let us examine the MFPT as a function of the stiffness $k$ of the potential for different locations $c$ of the reflective wall ($n$ being fixed). We consider two opposite limits: (i) When the wall is at $c=1$, \textit{i.e}, at the initial position of the particle, the particle cannot move to the right and only moves to the left. As $k$ increases, it has to climb a higher barrier to reach the target at 0 and hence the MFPT will increase monotonically with $k$. The stiffer the barrier, more time is needed to reach the target. (ii) The second is the opposite limit when $c\to \infty$, \textit{i.e}, the wall is far away to the right. In this case, when $k\to 0$ (free diffusion), the MFPT is infinite since there are trajectories that wander away to infinity in the direction opposite to the target. On the other hand, when $k\to \infty$ (extremely stiff potential centered around $x=1$), the particle gets fully localised at $x=1$ and is prevented to reach the target, hence the MFPT again diverges. Thus the MFPT, as a function of $k$ (for fixed $c\to \infty$) diverges at the two limits $k\to 0$ and $k\to \infty$. Consequently, it must have a minimum at some intermediate finite value $k=k_{\rm opt}$. As $c$ varies from 1 to $\infty$, the above reasoning suggests that there is a likely critical value $c^*$ at which the situation changes from (i) to (ii), {\it i.e.}, the MFPT vs. $k$ curve must change from a monotonically increasing [situation (i)] to a non-monotonic behavior with a minimum at $k_{\rm opt}$ [situation (ii)]. Therefore, at fixed shape parameter $n$, the optimal potential stiffness undergoes a continuous \lq\lq freezing\rq\rq\ transition with the domain size, where the optimal stiffness is the \lq\lq order parameter\rq\rq\ and $c$ the \lq\lq control parameter\rq\rq. 

\begin{figure}[htp]
\centering
			\includegraphics[width=\textwidth]{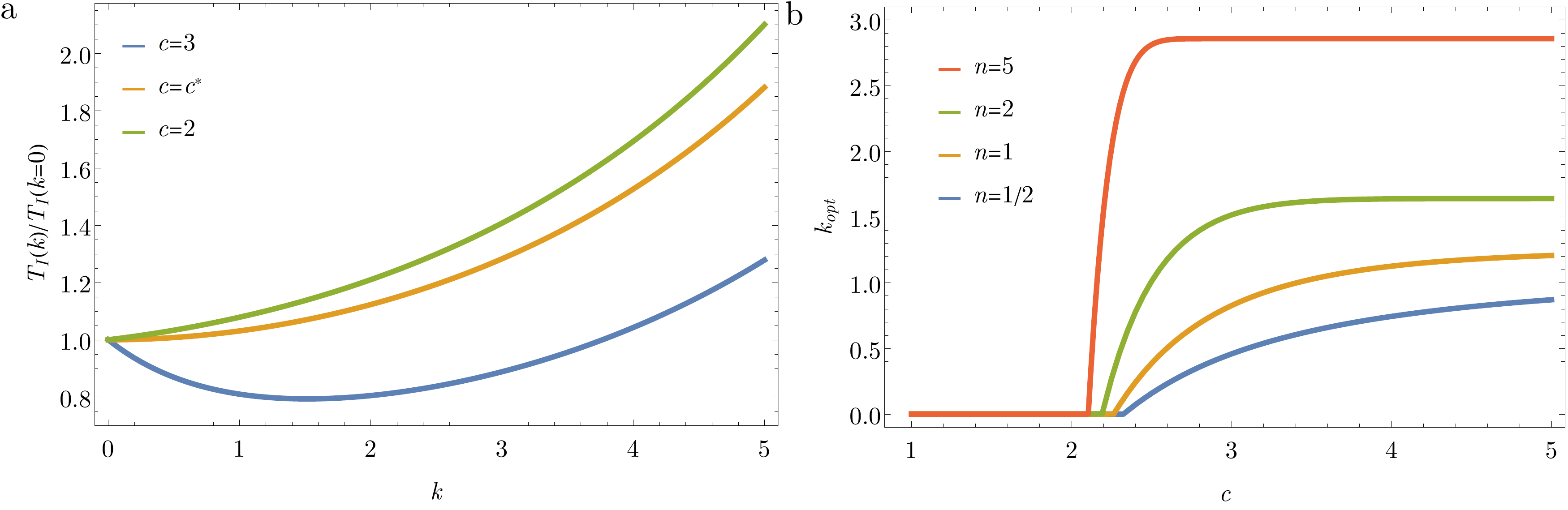}
			\caption{Searches starting from $x_0=1$.  (a) Normalized mean first passage time $T_I(k)/T_I(k=0)$ as a function of the potential stiffness for a harmonic potential ($n=2$) and different values of the domain size $c$. (b) Optimal potential stiffness $k_{\rm opt}$ as a function of the domain size $c$ for different potential exponent $n$.}
			\label{fig:fig2ab}
\end{figure}

 The analysis of Eq. (\ref{mfptVn}) confirms this scenario, as shown by Figs. \ref{fig:fig2ab}a and \ref{fig:fig2ab}b. The exact value of $c^*$ depends on the exponent $n$ and corresponds to the point at which the slope of $T_I$ at $k=0$ changes from positive to negative values, \textit{i.e.},
\begin{equation}
    \frac{\partial T_{I}(k,c^*)}{\partial k}\Bigg |_{k=0}=0.\label{Tpartialk}
\end{equation}

Taking the derivative of Eq. (\ref{mfptVn}) we get
\begin{equation}
\fl
        \frac{\partial T_I(k,c)}{\partial k}\Bigg |_{k=0}=\frac{1}{n}\int^1_{0}dy \int^{c}_ydz\ \left(|y-1|^n-|z-1|^n\right)=\frac{c-1-(c-1)^{n+1}+\frac{n}{n+2}}{n(n+1)}.\label{DmfptVn}
\end{equation}
Setting the above expression equal to zero we find
\begin{equation}
    c^*=1+\ell,\label{criticalc}
\end{equation}
where $\ell$ is the positive root of the equation
\begin{equation}
\ell^{n+1}-\ell-\frac{n}{n+2}=0.
\end{equation}
In the particular case $n=1$, this equation is easily solved as
\begin{equation}
c^*(n=1)= \frac{3 + \sqrt{7/3}}{2}=2.26376\dots
\end{equation}

\begin{figure}[htp]
\centering
			\includegraphics[width=\textwidth]{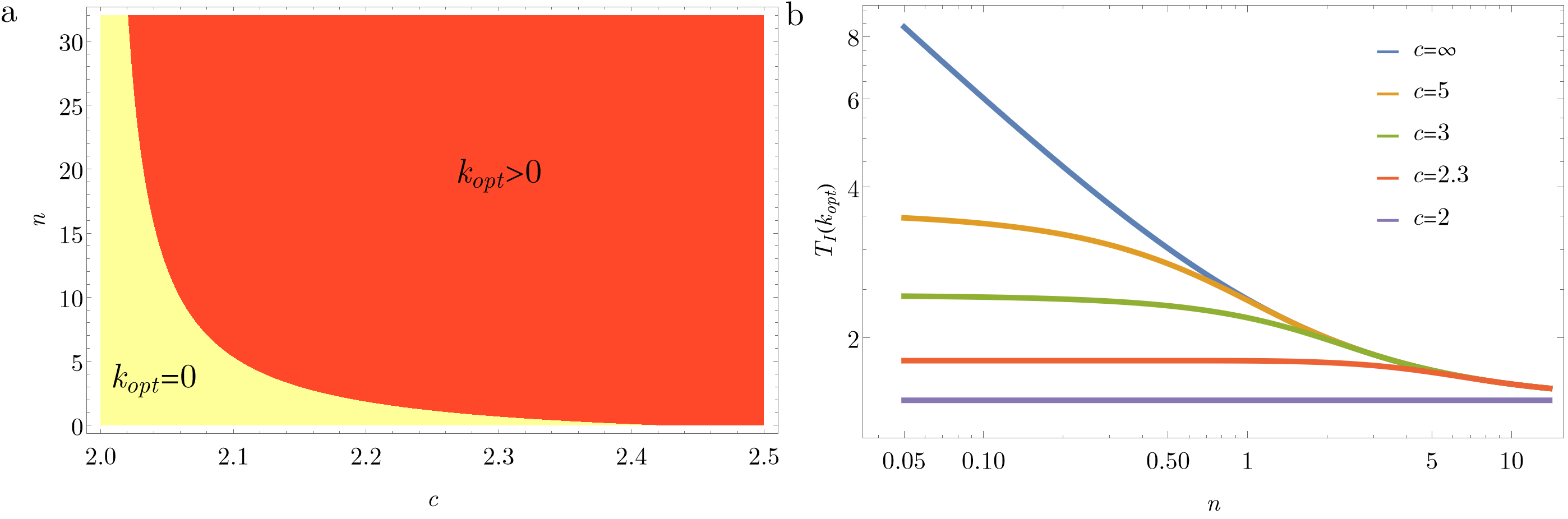}
			\caption{Searches starting from $x_0=1$. (a) Phase diagram in the ($c,n$)-plane. In the red area the MFPT is minimum at a potential stiffness $k_{\rm opt}>0$, whereas in the yellow region simple diffusion ($k=0$) is optimal. 
			(b) MFPT evaluated at the optimal potential stiffness $k_{\rm opt}$ as a function of $n$ for several values of $c$, including the case of the semi-infinite line $c=\infty$.}
			\label{fig:fig3ab}
\end{figure}

The value of the critical domain size $c^*$  decreases with $n$, as illustrated by the diagram \ref{fig:fig3ab}a showing the two dynamical phases. The value of $c^*$ is maximal when $n\to 0$, and minimal at $n=\infty$. In the limit $n\to\infty$, Eq. (\ref{criticalc}) gives $c^*(n\to\infty) \approx 2+\ln(2)/n+\mathrm{O}(1/n^2)$,  therefore, $c^*(n=\infty)=2$.  Conversely, in the limit $n\to 0$, the equation for the root $\ell$ simplifies to $ \ell \ln(\ell)=1/2 +\mathrm{O}(n)$, whose solution leads to the non-trivial value $c^*(n\to 0)= 2.42153...$.

\begin{figure}[htp]
\centering
			\includegraphics[width=.5\textwidth]{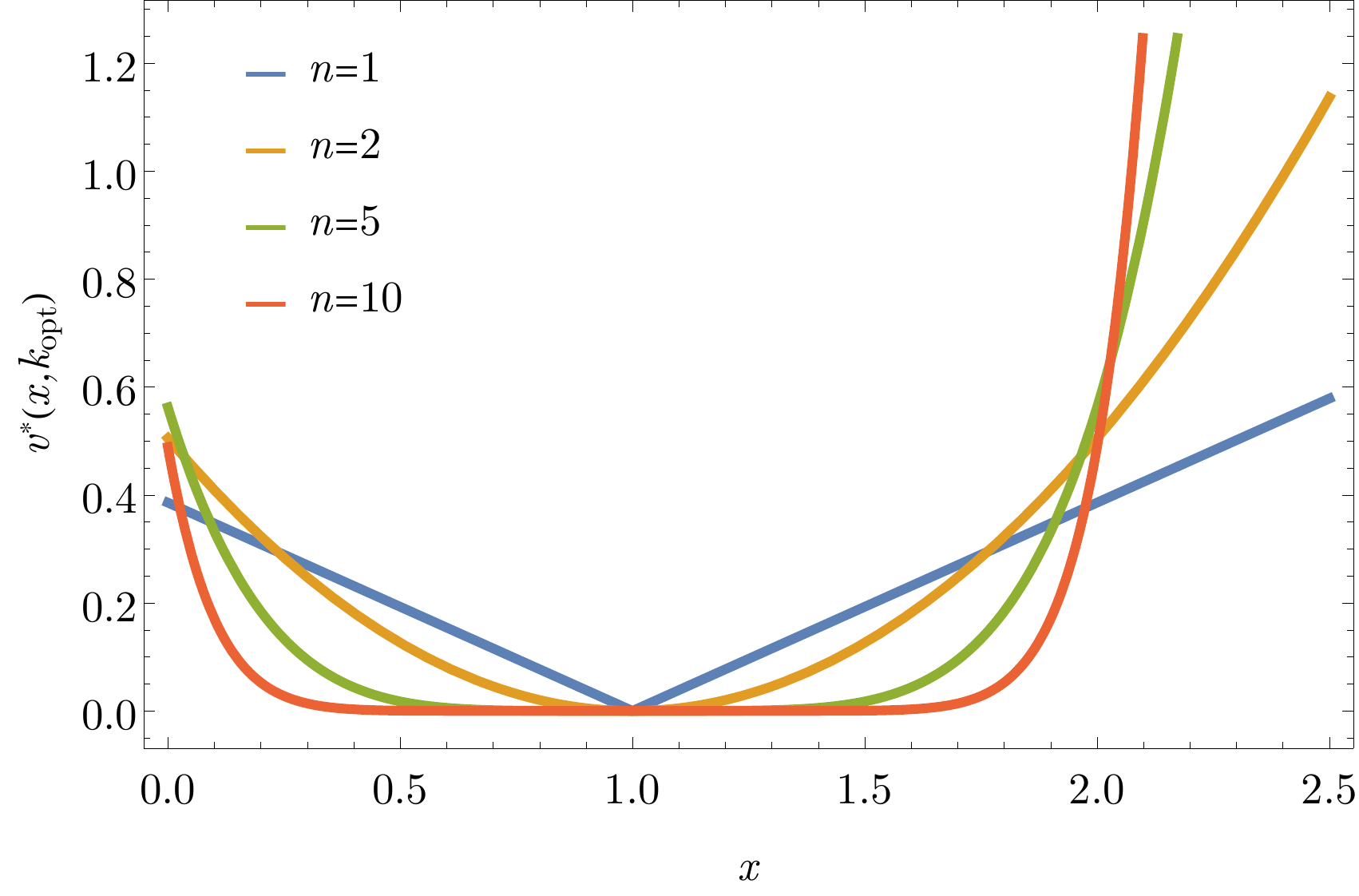}
			\caption{Potential $v(x)=k|x-1|^n/n$ as a function of $x$, evaluated at the optimal potential stiffness $k_{\rm opt}$ for different values of $n$. The domain size is fixed to $c=2.5$. }
			\label{fig:fig4}
\end{figure}

In Fig. \ref{fig:fig3ab}b, we observe that the optimal MFPT, or $T_{I}(k=k_{\rm opt})$, is a decreasing function of $n$ for any $c>2$. Therefore the best possible potential is the one corresponding to $n\to \infty$, or $v(x)=0$ for $x<2$ and $v(x)=\infty$ for $x>2$, which is equivalent to having a freely diffusing particle in a domain of size 2 (or $2X_0$ in the original coordinate). We show in Fig. \ref{fig:fig4} the shape of several optimal potentials $v^*(x)=k_{\rm opt}|x-1|^n/n$ for different values of the potential exponent $n$, in a domain of size $2.5$.    

\subsection{Comparison with optimal resetting}\label{secResetting}

The above results show the existence of a critical domain size $c^*$ below which the optimal stiffness $k_{\rm opt}$ freezes to the value 0. For completeness and to provide a coherent picture, we revisit in this section a problem where a similar transition was found in bounded domains, for a free Brownian particle subject to resetting to the initial position. Following  \cite{Evans_2011,EvansPRL2011,Pal2019}, let us consider a Brownian particle that diffuses in the interval $[0,c]$ (where $v(x)=0$) and is reset to its initial position at exponentially distributed times with mean $1/r$. The boundaries at $x=0$ and $c$ are absorbing and reflective, respectively, like in set-up I. Whereas diffusion in a static confining potential $v(x)$ describes an equilibrium process, resetting violates local detailed balance and generates dynamically a non-equilibrium steady state with an effective linear potential similar to the case $n=1$ and where $k$ is substituted by $\sqrt{r}$ \cite{Evans_2013}. Although the results shown below have already been developed previously under some form \cite{Saeed2019,Evans_2020}, we would like to emphasize that the MFPT in the resetting problem has a very different behaviour with respect to the domain size.

The Laplace transform of the survival probability of a process under stochastic resetting, $\widetilde{Q}_r(x,s)$, can be related to the same quantity for the process in the absence of resetting, $\widetilde{Q}_0(x,s)$, through the relation\cite{Evans_2020}
\begin{equation}
    \widetilde{Q}_r(x,s)=\frac{\widetilde{Q}_0(x,s+r)}{1-r\widetilde{Q}_0(x,s+r)},\label{Qres}
\end{equation}
where the resetting position coincides with the initial position $x$.
The survival probability $\widetilde{Q}_0(x,s)$ for a diffusive particle solves Eq. (\ref{SurvPs}) with the potential $v(x)=0$, and satisfies the boundary conditions $Q(x=0,t)=0$ and $\partial_x Q(x,t)|_{x=c}=0$. Hence $\widetilde{Q}_0(x,s)$ is given by
\begin{equation}
    \widetilde{Q}_0(x,s)=\frac{1}{s}-\frac{\cosh{\sqrt{s}(c-x)}}{s \cosh{\sqrt{s}c}}.\label{LQdif}
\end{equation}
Inserting the relation (\ref{LQdif}) into (\ref{Qres}) and using (\ref{relQT}), one obtains the MFPT under resetting, $T_r(x)$:
\begin{equation}
    T_r(x)=\frac{\cosh{\sqrt{r}c}}{r\cosh{\sqrt{r}(c-x)}}-\frac{1}{r}.\label{Tres}
\end{equation}
(It is implicit in the above expression that the resetting rate is expressed in units of $D/L^2$, the inverse of the characteristic time introduced at the beginning of section \ref{generalsetup}.)

\begin{figure}[htp]
\centering
			\includegraphics[width=\textwidth]{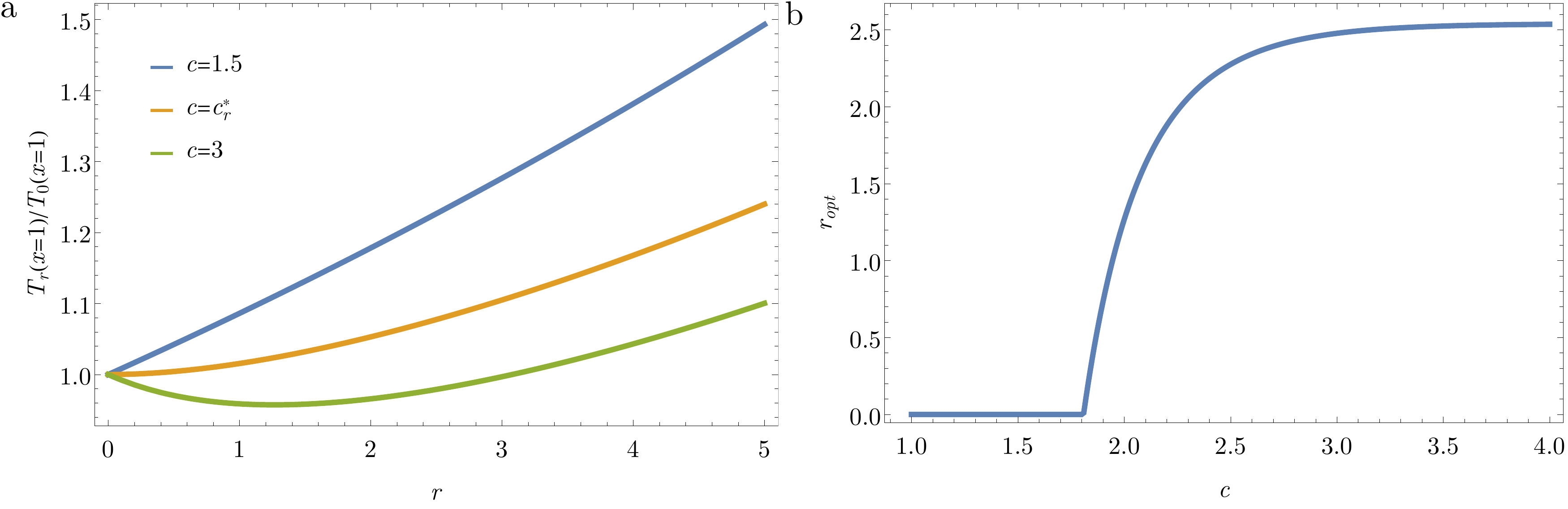}
			\caption{Brownian searches starting from $x=1$ and subject to stochastic resetting, in set-up I. (a) Mean first passage time as a function of the re-scaled resetting rate $r$ for different values of the domain size $c$. (b) Optimal resetting rate $r_{\rm opt}$ as a function of the domain size.}
			\label{fig:fig5ab}
\end{figure}

Fixing the resetting position to $x=1$, we see in Fig. \ref{fig:fig5ab}a-b that $T_r$ exhibits a behaviour with respect to $r$ similar to that of $T_I$ with $k$. When the domain size is below a critical value $c^*_r$, resetting does not improve target search and the minimum of the MFPT is reached at $r=0$ (free diffusion), whereas for $c\ge c^*_r$ the MFPT is optimized at a resetting rate $r_{\rm opt}$ which varies continuously from $0$. The critical size $c^*_r$ is determined from the condition
 \begin{equation}
    \frac{\partial T_r(r,c^*_r)}{\partial r}\Bigg |_{r=0}=0,\label{Trpartialk}
\end{equation}
which gives, from Eq. (\ref{Tres}),
\begin{equation}
    c^*_r=\frac{1}{1-\frac{1}{\sqrt{5}}}=1.80902\dots<2\label{cricx1}
\end{equation}
a result previously derived in \cite{Saeed2019}. In dimensional units, the critical domain size reads
\begin{equation}
    C_r^*=\frac{X_0}{1-\frac{1}{\sqrt{5}}}.
\end{equation}

\begin{figure}[htp]
\centering
			\includegraphics[width=.5\textwidth]{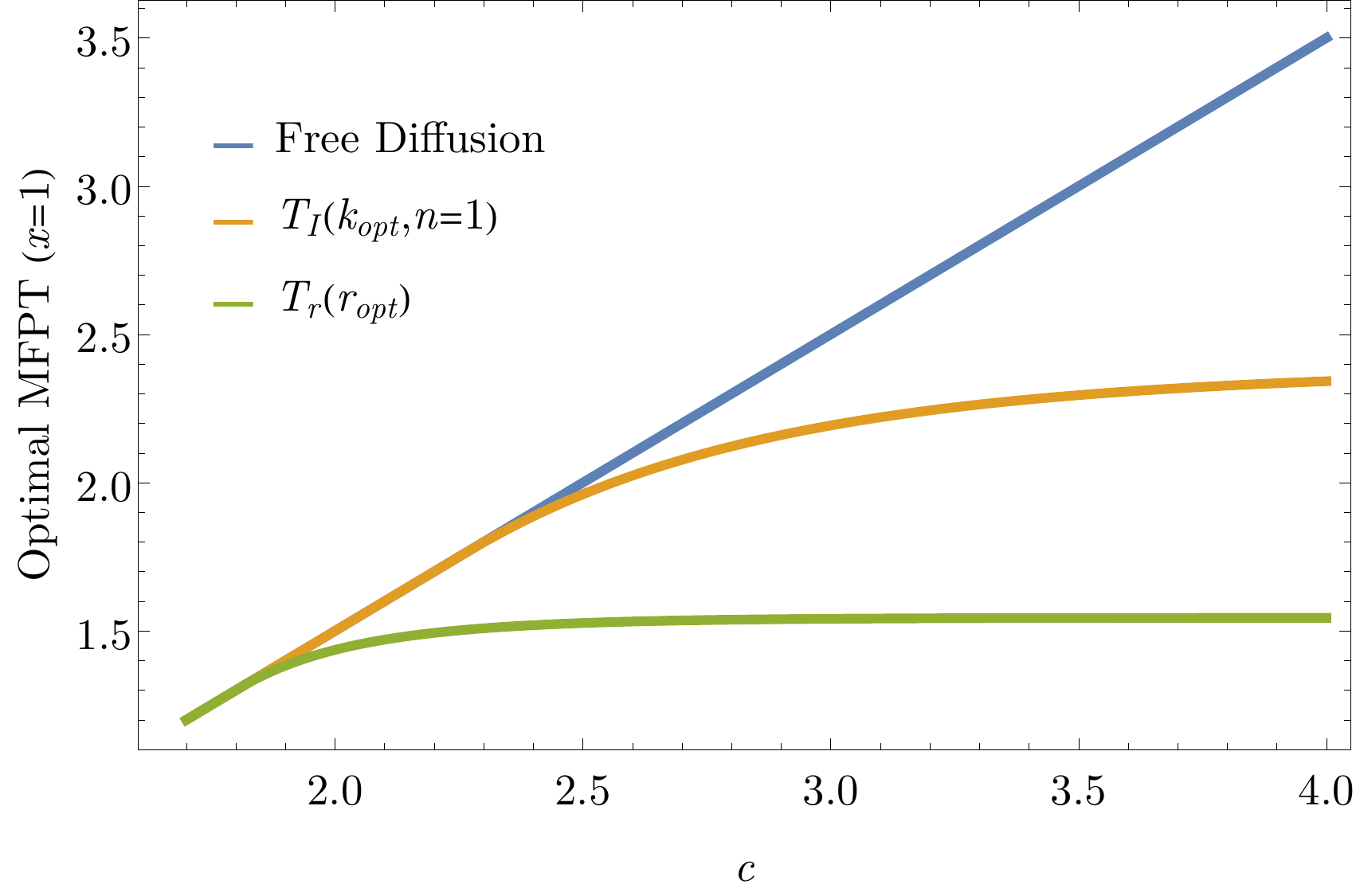}
			\caption{Optimal mean first passage time as a function of the domain size for a diffusive particle in an external potential $v(x)=k|x-1|$ with $k=k_{\rm opt}(c)$, or subject to stochastic resetting to the initial position $x=1$, at the optimal resetting rate $r_{\rm opt}(c)$. The free diffusion case also in shown.}
			\label{fig:fig6}
\end{figure}

The value of $c_r^*$ is lower than the critical domain size $c^*$ in the equilibrium case with any confining potential $v(x)=k|x-1|^n/n$. Hence, resetting has a negative effect on the MFPT over a smaller interval of domain sizes. In addition, as shown by Fig. \ref{fig:fig6}, searches at the optimal resetting rate $r_{\rm opt}$ are faster (for $c$ above $c_r^*$) than their counterparts that use an optimal potential with $n=1$, or $v^*(x)=k_{\rm opt}|x-1|$.  This result agrees with the unbounded domain case without the
reflective wall\cite{Evans_2013}. We also display for comparison the case of simple diffusion ($v(x)=0$). 

\section{Set-up II: absorbing/absorbing boundaries}\label{twotargets}

We now turn to the case of two absorbing boundaries. For convenience we now set the characteristic length $L$ equal to the domain size $C$. Hence, $c=1$ and the dimensionless space variable is $x=X/C$, which is comprised in the interval $[0,1]$. Again, we consider a particle with initial position at the minimum of the potential $(x=x_0)$, which can be varied in $[0,1]$.

Setting $c=1$ in Eq. (\ref{generalmfptVn}) and (\ref{A2n}) we obtain
\begin{equation}
\fl
        T_{II}(x_0)=\frac{\int^1_{0}dy \int^y_0dz\ e^{\frac{k}{n}\left(|y-x_0|^n-|z-x_0|^n\right)}}{\int^1_0 dy\ e^{\frac{k}{n}|y-x_0|^n}} \int^{x_0}_0 dy\ e^{\frac{k}{n}|y-x_0|^n} -\int^{x_0}_{0}dy \int^y_0dz\ e^{\frac{k}{n}\left(|y-x_0|^n-|z-x_0|^n\right)}.\label{mfptVn2}
\end{equation}
We investigate how the position $x_0$ affects the behavior of the MFPT. 

\subsection{Optimal potential stiffness  \texorpdfstring{$k$}{Lg}} 

In set-up I, we have seen that above a critical value of the domain size, the potential expedites target encounter as the stiffness $k$ increases from  0. Since the domain size was given in units of $x_0$ (the distance between the potential minimum and the target located at the origin), the above conclusion is equivalent to say that, if the domain size is fixed to unity, the presence of the confining potential can accelerate target encounter if the distance $x_0$ is smaller than $x_c=1/c^*$. If $x_0>x_c$, on the contrary, a weak potential delays search. Here, one would expect a similar behaviour if one sets $x_0$ close to any of the two absorbing boundaries. There should exist a $x_c$ such that, in the intervals $0<x_0<x_c$ and $1-x_c<x_0<1$, the MFPT decreases with $k$ at small $k$, whereas it increases in $[x_c,1-x_c]$. A similar situation is actually encountered (with respect to $r$) in the resetting problem with two absorbing boundaries \cite{Pal2019}. 

\begin{figure}[htp]
\centering
			\includegraphics[width=\textwidth]{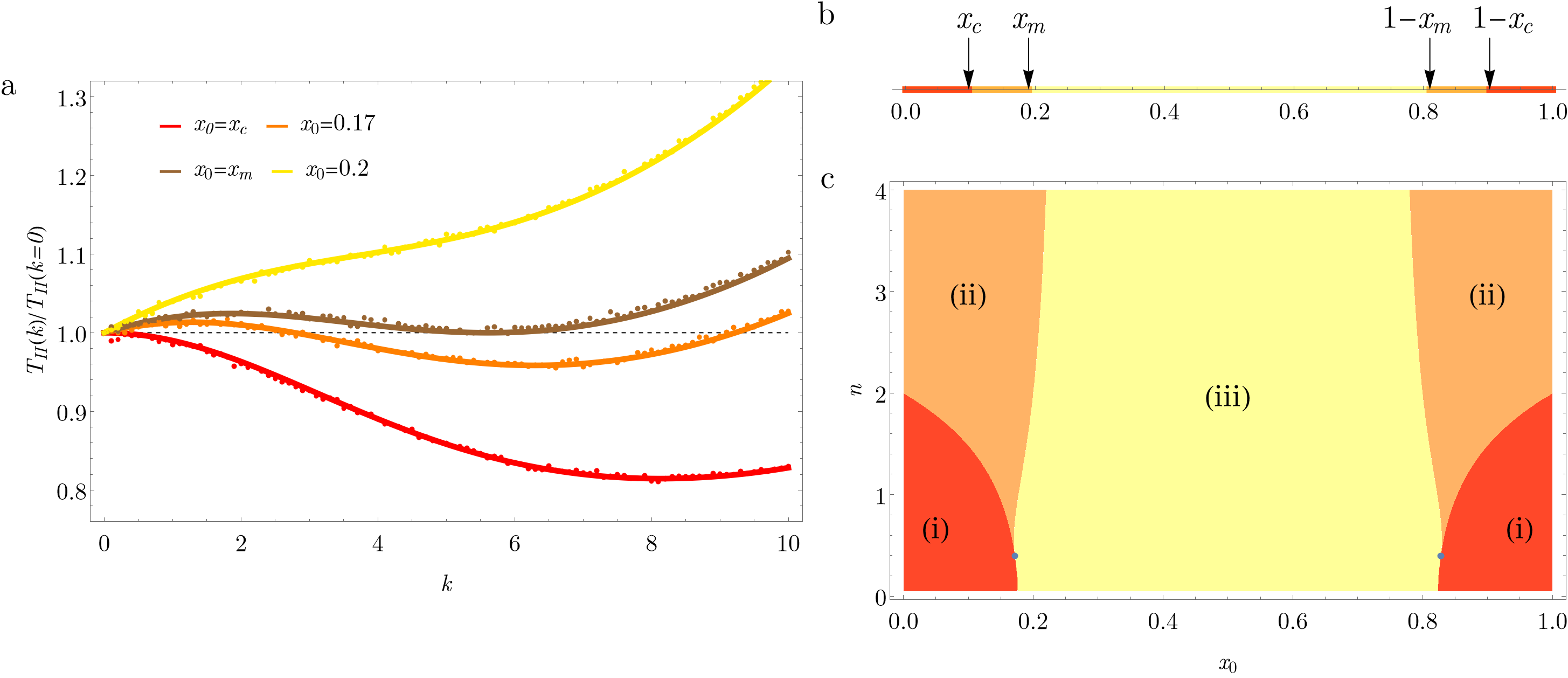}
			\caption{Mean exit time of a Brownian particle in set-up II, with a potential of the form $v(x)=k|x-x_0|^n/n$. The initial position of the particle is the potential minimum $x_0$.  (a) Ratio $T_{II}(k)/T_{II}(k=0)$ as a function of the potential stiffness $k$ in the case $n=1$, for different $x_0$ varied through the critical values  $x_c\simeq 0.146447$ and $x_m\simeq 0.177311$. The latter position defines a discontinuous transition for the optimal parameter, from a value $k_{\rm opt}\simeq5.61157$ to $k_{\rm opt}=0$. Symbols represents Monte Carlo simulations; each point is an average over $2\times10^5$ realizations. (b) Position of the transitions points $x_c$, $x_m$ and their symmetrical $1-x_m$, $1-x_c$, in the case $n=1.5$. (c) Phase diagram in the $(x_0,n)$-plane. In region (i) in red, $dT_{II}/dk|_{k=0}\le0$ and the MFPT gets minimized at $k_{\rm opt}>0$. In region (ii) in orange, $dT_{II}/dk|_{k=0}>0$ but the absolute minimum $k_{\rm opt}$ is still $>0$ (metastable region). In region (iii) in yellow, simple diffusion becomes optimal ($k_{\rm opt}=0$). The three phases meet at the triple point $(0.17185,0.39539)$ and its symmetrical  $(0.82814,0.39539)$
			}
			\label{fig:fig7ab}
\end{figure}

The behavior of Eq. (\ref{mfptVn2}) with $x_0$ is actually quite different from the single target case and the scenario sketched above. Due to the symmetry of the problem, let us focus on the case $x_0\in[0,1/2]$. There generally exist not two, but three distinct regions in this interval for the MFPT (see Fig. \ref{fig:fig7ab}a): 




(i) When $0<x_0<x_c$, the derivative $\partial_k T_{II}|_{k=0}$ is $<0$ and the equation $\partial_k T_{II}(k)=0$ has a root $k^*>0$, where $T_{II}$ reaches its absolute minimum. Therefore $k_{\rm opt}=k^*$.

(ii) When $x_c< x_0< x_m$, the derivative $\partial_k T_{II}|_{k=0}$ is $>0$, but $T_{II}(k)$ vs. $k$ has two local minima: one at $k=0$ and the other at $k=k^*$ where $\partial_k T_{II}(k^*)=0$.
In this region, $T_{II}(k^*)<T_{II}(0)$, thus the optimal parameter is still $k_{\rm opt}=k^*$, while $k=0$ is metastable.

(iii) When $x_0>x_m$, $T_{II}(0)$ is the absolute minimum ($k_{\rm opt}=0$). Thus
$x_m$ can be deduced by solving $T_{II}(k^*)= T_{II}(0)$.

Figure\ref{fig:fig7ab}b displays, for the particular case $n=1.5$, how the interval $[0,1]$ is subdivided into five regions where the three phases above are observed. Fig. \ref{fig:fig7ab}c shows the whole phase diagram in the $(x_0,n)$-plane, which we discuss qualitatively. The critical line $x_c(n)$ that limits the first region is obtained analytically (see below), whereas the line $x_m(n)$ is obtained from numerical minimization of Eq. (\ref{mfptVn2}). Surprisingly, when $n\ge2$, $x_c=0$, namely, the region (i) disappears. This means that for harmonic or higher order potentials, the MFPT always increases with $k$ at small $k$, for any values of $x_0$ in the interval. 

Notably, when the starting position $x_0$ crosses the boundary from region (ii) to (iii) at $n$ fixed, the optimal stiffness $k_{\rm opt}$ undergoes a discontinuous "first order" transition, with an abrupt jump from the finite $k^*$ to $0$ (see Fig. \ref{fig:fig8ab}b further). 

At small enough $n$, the region (ii) is no longer present: one then goes directly from region (i) to (iii) through a continuous decrease of the optimal stiffness $k^*$ to $0$. The intersection of $x_c(n)$ and $x_m(n)$ thus defines a tricritical point ($x_t$,$n_t$), which we determine in the following. 

\subsubsection{Transition line  \texorpdfstring{$x_c(n)$}{Lg}}
\hfill\\

This position, when it exists, is defined by the equation
\begin{equation}
    \frac{\partial T_{II}(k,x_c)}{\partial k}\Big |_{k=0}=0.\label{Tpartialk2}
\end{equation}
Taking the derivative of Eq. (\ref{mfptVn2}) and evaluating at $k=0$ we obtain
\begin{eqnarray}
\fl
        \frac{\partial T_{II}(k,x_c)}{\partial k}\Bigg |_{k=0}=&\frac{1}{n}\frac{\int^1_{0}dy \int^y_0dz }{\int^1_0 dy} \int^{x_c}_0 dy\ |y-x_c|^n +\frac{1}{n}\frac{\int^1_{0}dy \int^y_0dz\ \left(|y-x_c|^n-|z-x_c|^n\right) }{\int^1_0 dy\ } \int^{x_c}_0 dy\\
        &-\frac{1}{n}\frac{\int^1_{0}dy \int^y_0dz}{\left(\int^1_0 dy\right)^2} \int^{x_c}_0 dy \int^1_0 dy\ |y-x_c|^n-\frac{1}{n}\int^{x_c}_{0}dy \int^y_0dz\ \left(|y-x_c|^n-|z-x_c|^n\right),\nonumber\label{DmfptVn2}
\end{eqnarray}
which gives the following equation for $x_c$:
\begin{equation}
   (n+2-4x_c)x_c^{n}+(n-2+4x_c)(1-x_c)^{n}=0.\label{rootx0}
\end{equation}
The numerical resolution of Eq. (\ref{rootx0}) shows that when $0<n<2$, there exist two solutions in the interval $[0,1]$: $x_c$ and the symmetrical $1-x_c$. For $n>2$ the real solutions fall outside $[0,1]$ and become unacceptable. It is remarkable that the transition between these two behaviors occurs for the generic harmonic potential $n=2$, where Eq. (\ref{rootx0}) is easily solved, yielding $x_c=0$ and 1. 

In the limit $n\to0$ (corresponding to the logarithmic potential), one finds the non-trivial roots $x_c=0.176041...$ and  $1-x_c=0.823959...$ .

\subsubsection{Triple point}
\hfill\\

At the tricritical point, $x_c(n)$ and $x_m(n)$ intersect and the phases (i), (ii) and (iii) coexist. Hence, the condition (\ref{Tpartialk2}) or (\ref{rootx0}) for $x_c(n)$ must be met. A second condition is deduced by noticing that, fixing $n$ and varying $x_0$, the transitions (ii)$\rightarrow$(iii) and (i)$\rightarrow$(iii) differ in the concavity of the MFPT at small $k$, which implies that $\partial^2_k T_{II}|_{k=0}=0$ right at the triple point separating them.
More precisely, let us consider a point $(x_0,n)$ in the vicinity of $(x_t,n_t)$
and expand $T_{II}$ in powers of $k$:
\begin{equation}
T_{II}(k)=T_0+ak+bk^2+dk^3+...,
\end{equation}
with $d>0$.
In the metastable region (ii), $a>0$ and as $k$ increases from 0 the MFPT has a local maximum before reaching its global minimum. This implies that $b<0$. Across the transition (i)$\rightarrow$ (iii), on the other hand, $a$ changes sign. Let us assume that $b<0$ in this case, too, and show that this leads to a contradiction.
Consider a point in the region (iii) very close to the phase (i), {\it i.e.}, $a>0$ but arbitrarily small. Solving 
$\partial_k T_{II}(k)=0$ gives the local minimum $k^*\simeq-\frac{2b}{3d}+\frac{a}{2b}$ and $T(k^*)\simeq T_0+\frac{20 b}{27d^2}-2ab\simeq T_0+\frac{20 b}{27d^2}<T_0$ since $b<0$. Hence $T_0$ is not the absolute minimum, which contradicts the definition of phase (iii). Therefore $b$ is positive across the transition (i)$\rightarrow$ (iii). By continuity, $b$ must vanish at the triple point, or
\begin{equation}
    \frac{\partial^2 T_{II}(k,x_t)}{\partial k^2}\Big|_{k=0}=0.
\end{equation}
The tricritical point ($x_t$,$n_t$) is thus a point of inflection where both $\partial_k T_{II}$ and $\partial^2_k T_{II}$ vanish at $k=0$. 
The second condition is obtained from deriving Eq. (\ref{mfptVn2}) twice. After a lengthy calculation, one obtains
\begin{eqnarray}
\fl
    \frac{\partial^2 T_{II}}{\partial k^2}\Big|_{k=0}=&\frac{x_0
   \left(1-x_0\right){}^{2 n+1}}{(n+1)^2 (2 n+1)}+\frac{x_0 \left(1-x_0\right)\left(x_0^n-\left(1-x_0\right)^n\right)}{n^2}  \Big(\frac{x_0^n+\left(1-x_0\right)^n}{2 (2
   n+1)}\nonumber\\
   &-\frac{4 x_0^2
   \left(x_0^n-\left(1-x_0\right){}^n\right)-(n-2) \left(1-x_0\right){}^n}{(n+1)^2 (n+2)}\nonumber\\
   &-\frac{2 x_0 \left[(n+2-4x_0) x_0^n+(n-2+4 x_0) \left(1-x_0\right)^n\right]}{(n+1)^2 (n+2)}\Big).
\end{eqnarray}
Setting the above expression to zero and combining it with Eq. (\ref{rootx0}) boils down to the rather simple system 
\begin{eqnarray}
(n+2-4x_0)x_0^{n}+(n-2+4x_0)(1-x_0)^{n}=0,\\
    (n+2) [(n-4) n+2]+16 (n-1)\left(1-x_0\right) x_0=0,
\end{eqnarray}
whose numerical solution in the interval $x_t<1/2$ is
\begin{eqnarray}
   n_t=0.395391...\\ x_t=0.171859... 
\end{eqnarray}
The second triple point is at $(n_t,1-x_t)$ by symmetry.

\subsection{Comparison with optimal resetting}

The case $n=1$, which is the closest to diffusion with stochastic resetting, exhibits four transitions as $x_0$ is varied in the phase diagram \ref{fig:fig7ab}b. In the resetting problem with the same set-up, however, there are only two transition points, as the metastable behaviour leading to region (ii) is absent. With two absorbing boundaries,
the mean exit time $T_r$ of a Brownian particle subject to resetting to the initial position can be calculated following the same route as in section \ref{secResetting} with $c=1$ (see also \cite{Pal2019}):
\begin{equation}\label{resetsetupII}
    T_r=\frac{1}{r}\left(\frac{\sinh{\sqrt{r}}}{\sinh{(1-x_0)\sqrt{r}}+\sinh{x_0\sqrt{r}}}-1\right).
\end{equation}
When the resetting rate $r$ is varied at fixed $x_0$, the MFPT above exhibits only one local minimum, at $r=0$ or at a finite value. The transition occurs when $\partial_r T_r|_{r=0}=0$, which gives $x_{c,r}=(5-\sqrt{5})/10=0.27639...$ 
(symmetrically the second transition occurs at $1-x_{c,r}=(5+\sqrt{5})/10$) \cite{Pal2019}.
\begin{figure}[hpt]
\centering
			\includegraphics[width=\textwidth]{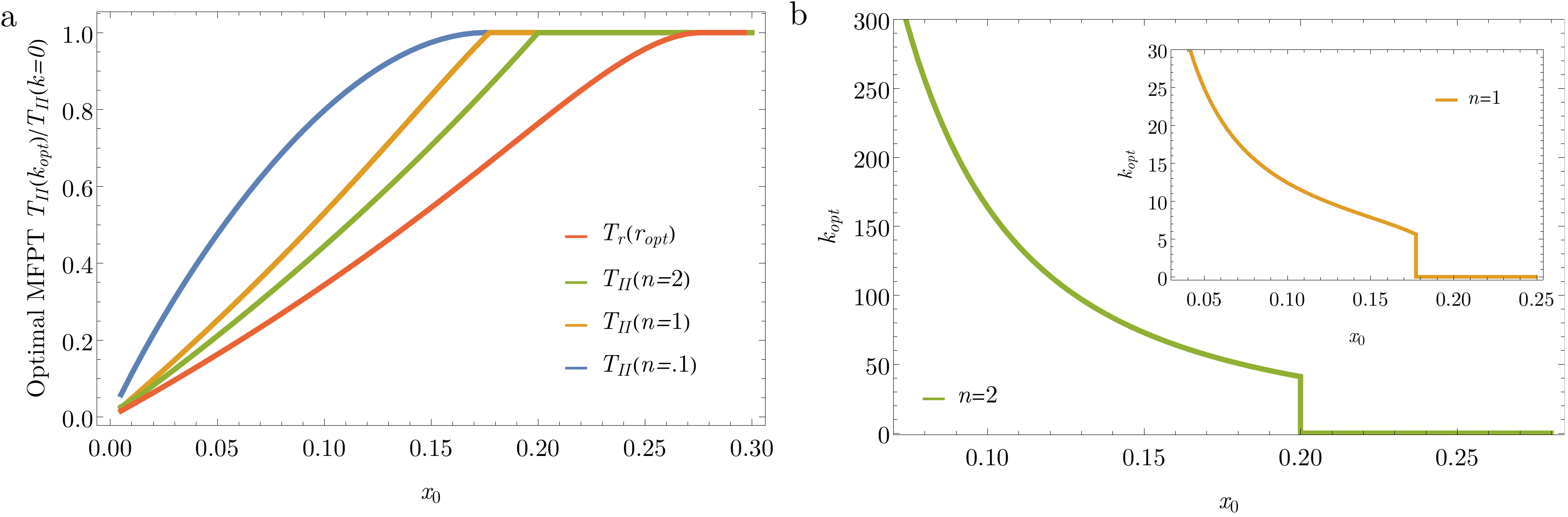}
			\caption{(a) Mean absorption time of a Brownian particle in set-up II in the presence of an external potential of the form $v(x)=k|x-x_0|^n/n$ at $k=k_{\rm opt}$, or under the action of stochastic resetting at $r=r_{\rm opt}$ (red curve). Searches start from the resetting/potential minimum position $x_0$. The times are normalized by the free diffusion MFPT for each $x_0$. (b) Optimal stiffness as a function of $x_0$. In the harmonic case $n=2$, the discontinuous transition occurs at $x_m=0.199951...$, and for $n=1$, at $x_m=0.177311...$.}
			\label{fig:fig8ab}
\end{figure}

Fig. \ref{fig:fig8ab}a depicts the optimal MFPTs obtained by  minimizing Eq. (\ref{resetsetupII}) with respect to $r$ at fixed $x_0$, as well as the ones corresponding to several optimal potentials. These times are normalized by the free-diffusion
MFPT ($r=0$ or $k=0$) for each $x_0$. Once again, optimal resetting is always more efficient than the optimal equilibrium dynamics. Both processes, however, become much more efficient than free diffusion as $x_0$ gets closer to one of the targets.


Further analysis of Eq. (\ref{resetsetupII}) shows that $\partial^2_rT_r|_{r=0}>0$, which implies that the freezing transition between the regimes $r_{\rm opt}=0$ and $r_{\rm opt}>0$ is continuous, in contrast with the discontinuous behavior of $k_{\rm opt}$ in the potential case when $n>n_t$. This is confirmed by Fig. \ref{fig:fig8ab}b, where the optimal parameter $k_{\rm opt}$ for $n=1$ or 2 jumps from a non-zero value to $0$ when $x_0$ crosses $x_m$ from below.

\section{Discussion}\label{discussion}

We have computed the mean first passage time of a one-dimensional Brownian particle to an absorbing boundary in the interval $[0,c]$, in the presence of an external confining potential of the form $v(x)=k|x-x_0|^n/n$, where $x_0\in[0,c]$ is also the starting position. When an absorbing boundary is placed at the origin and a reflective wall at $c$, the search time can be minimized by a suitable choice of the potential stiffness $k$. The optimal stiffness $k_{\rm opt}$ undergoes a second order "freezing" transition as the relative distance $c/x_0$ crosses a critical number $c^*(n)$, which depends on $n$. When  $c/x_0 > c^*(n)$, the additional confinement produced by the potential can help complete the search process faster, or $k_{\rm opt}>0$. Otherwise, search is delayed by the potential, or $k_{\rm opt}=0$. 

This freezing transition is analogous to the one observed in a non-equilibrium situation, {\it i.e.}, a diffusive process with stochastic resetting to the starting position. In this case the resetting rate $r$ plays a role analogous to $k^2$ in the case $n=1$. Below a relative domain size $c/x_0=c_r^*$, stochastic resetting is no longer beneficial either compared to simple diffusion, but the critical number $c_r^*$ is smaller than $c^*(n)$ for all $n$. Hence, compared to any optimal equilibrium situation, optimal resetting processes not only allow faster target encounters but they are also useful in smaller domains.

When both ends of the interval are absorbing, the optimization of the MFPT is more intricate. Setting $c=1$ for convenience, we found the existence of three different "phases" in the $(x_0,n)$-plane for which the behaviour of the MFPT is completely different. In the phase (i), the MFPT decreases with increasing $k$ from zero and reaches a minimum at a non-zero potential stiffness. In the phase (ii), the MFPT is metastable: it first increases with $k$, goes through a local maximum and then decreases to reach a global minimum at $k_{\rm opt}>0$, like in phase (i). In the phase (iii),  pure diffusion is optimal ($k_{\rm opt}=0$). For the transition (i) $\to$ (iii) the optimal stiffness undergoes a continuous transition. In contrast, the transition (ii)$\to$(iii) is discontinuous. Somehow unexpectedly, region (i) disappears for potentials with  $n\ge 2$: in other words, the additional confinement produced by a weak harmonic (or higher order) potential will always delay absorption compared to simple diffusion, independently of the starting position/domain size.

Notably, there are two symmetrical tricritical points in the ($x_0,n$)-plane, at which the three above phases coexist: for a potential exponent $n<n_t=0.395391...$, the system can only be in the phases (i) and (iii). For $n=1$, which is above $n_t$, there are four transitions as $x_0$ is varied. This behaviour of the MFPT is not observed for a diffusive process with stochastic resetting to the origin, where the transitions in the optimal resetting rate are continuous and where the absence of metastability effects limits the number of dynamical phases to two. 

The study of optimal searches under stochastic resetting to the initial position is  simplified by the existence of a universal criterion, valid for any background process \cite{ReuveniPRLoptimal,PalPRL2017,BelanPRL2018,Pal2019,ray2019peclet}. The criterion tells that if the coefficient of variation ($CV$) of the first passage times of the underlying process (without resetting) is larger than unity, then resetting will produce a speed-up of the search process. Otherwise, resetting will be detrimental. Hence, the transition points $c_r^*$ (in set-up I) or $x_0=x_{c,r}$ (in set-up II) are such that the $CV$ corresponding to free Brownian motion is exactly 1. Similarly, it would be useful to predict whether an external potential can decrease or not the mean search time, based only on a property of the process without potential. However, no such universal relation exists in this case. This can be seen by noticing that the transition points $c^*(n)$ or $x_c(n)$ depend on $n$, {\it i.e.}, on the shape of the potential that is going to be applied. As $c^*(n)$ varies between 2 and $2.42153...$ in set-up I (see also Fig. \ref{fig:fig3ab}), it is unlikely that a Brownian quantity related to a first passage time remains constant in this range. However we can make a few general statements: as mentioned above, with two absorbing boundaries, harmonic potentials of weak stiffness cannot improve the mean absorption time, while weak piece-wise linear potentials can. 

Our study also shows that one should be cautious in drawing conclusions about the negative effects on the MFPT of an increasing potential strength (or resetting rate), based on a perturbative analysis of the free diffusion case. Such analysis, which leads to the general criteria $CV=1$ mentioned above for resetting processes, does not guarantee that an absolute minimum does not exist for a large value of the parameter considered, out of the regime of validity of the calculation. Metastable minima of the MFPT analogous to the ones found here by varying $k$ were recently observed with respect to the resetting rate for a Brownian particle in a bounded domain and subject to different potential shapes \cite{ahmad2022first}. In that study, the number of absorbing boundaries also played a crucial role on the order of the transition for the optimal resetting rate. With two absorbing walls, tricritical points were usually present. Similar transitions have been found in drift-diffusion processes and Michaelis-Menten chemical reactions with restart \cite{pal2019landau}.

Metastability in the mean first passage time was also observed experimentally and theoretically for a free Brownian particle under resetting to a distribution of resetting positions, given by the Boltzmann-Gibbs distribution in a harmonic potential\cite{besga2020optimal,besga2021dynamical,Faisant_2021}. When the distance to the target is of the order of the width of this distribution, the MFPT exhibits a metastable behavior as the resetting periodicity is varied. These results were further generalized to the case of a one-dimensional persistent random walks in a harmonic potential, for which the distribution of resetting positions is non-Boltzmann\cite{tucci2022first}. Metastability is also the outcome of other protocols that use intermittent potentials to mimic non-ideal resetting processes\cite{Mercado_V_squez_2020}. 

To summarize, we have studied under which circumstances the application of an external confining potential acting on a Brownian particle in a bounded domain enhances target encounter, or when it is more convenient to let the particle diffuse without any external forces. Our analysis reveals some intricate effects of the potential shape on the optimal search strategy. This calls for seeking new optimal protocols that employ confining potentials in bounded or infinite domains, for instance potentials that fluctuate in time \cite{ResonantDoering1992,ResonantStochastic2021,Mercado_V_squez_2020}.

\ack 
GMV thanks CONACYT (Mexico) for a scholarship support. We acknowledge support from Ciencia de Frontera 2019 (CONACYT), project \lq\lq Sistemas complejos estoc\'asticos: Agentes m\'oviles, difusi\'on de part\'iculas, y din\'amica de espines\rq\rq \ (Grant 10872).

\section*{References}
\bibliographystyle{iopart-num}
\bibliography{Biblio}


\end{document}